\documentclass{iau} 
\usepackage{graphicx}

\title[Testing the robustness of MBH] 
{Testing the robustness of black 
hole mass measurements with 
ALMA and MUSE}

\author[Thater et al.]   
{Sabine Thater$^{1,2}$,
Davor Krajnovi\'{c}$^2$, Dieu D. Nguyen$^3$, Satoru Iguchi$^3$, Peter M. Weilbacher$^2$}

\affiliation{$^1$Department of Astrophysics, University of Vienna,\\ T\"urkenschanzstrasse 17, 1180 Wien, Austria\\ email: {\tt sabine.thater@univie.ac.at} \\[\affilskip]
$^2$Leibniz-Institute for Astrophysics Potsdam (AIP),\\An der Sternwarte 16, 14482 Potsdam, Germany\\[\affilskip]
$^3$National Astronomical Observatory of Japan,\\2 Chome-21-1 Osawa, Mitaka, Tokyo 181-0015, Japan\\[\affilskip]}

\pubyear{2019}
\volume{353}  
\jname{Galactic Dynamics in the Era of Large Surveys}
\editors{M. Valluri, \& J. A. Sellwood  }
\begin{document}

\maketitle

\begin{abstract}
We present our ongoing work of using two independent tracers to estimate the supermassive black hole mass in the nearby early-type galaxy NGC 6958; namely integrated stellar and molecular gas kinematics. We used data from the Atacama Large Millimeter/submillimeter Array (ALMA), and the adaptive-optics assisted Multi-Unit Spectroscopic Explorer (MUSE) and constructed state-of-the-art dynamical models. The different methods provide black hole masses of $(2.89\pm 2.05) \times 10^8M_{\odot}$ from stellar kinematics and $(1.35\pm 0.09) \times 10^8M_{\odot}$ from molecular gas kinematics which are consistent within their $3\sigma$ uncertainties. Compared to recent M$_{\rm BH}$ - $\sigma_{\rm e}$ scaling relations, we derive a slightly over-massive black hole. Our results also confirm previous findings that gas-based methods tend to provide lower black hole masses than stellar-based methods. More black hole mass measurements and an extensive analysis of the method-dependent systematics are needed in the future to understand this noticeable discrepancy.

\end{abstract}

\firstsection 
\section{Introduction}
Vast improvements in astronomical instrumentation have led to more than hundred robust massive black hole mass ($M_{\rm BH}$) measurements in local galaxies (e.g., \cite[van den Bosch 2016]{Bosch2016}, \cite[Saglia et al. 2016]{Saglia2016}). Combining these mass estimates with different host galaxy bulge properties, such as bulge mass, stellar velocity dispersion, and light concentration, revealed remarkably tight correlations which suggest a co-evolution between the massive black hole and their host galaxy (\cite[Kormendy \& Ho 2013]{Kormendy2013}). 
However, determining $M_{\rm BH}$ is a challenging procedure, and it is not possible to use one single method across the full sample of galaxies. The methods differ in their model assumptions and in the dynamical tracer (e.g., stars or gas) which is used to probe the gravitational potential in the vicinity of the massive black hole. Problematically, measurements from different dynamical tracers often give discrepant results, raising the question whether the variety of methods forces an additional bias on the scaling relations. Connecting mass results from different methods is necessary to evaluate the robustness and universality of the measurement results and thus crucial for improving the understanding of the interplay between the central black holes and their host galaxies. In this work, we make use of state-of-the-art instruments and modeling methods to add a new mass comparison to the catalog and investigate possible measurement systematics.

\section{NGC 6958 - an ideal candidate}
Different $M_{\rm BH}$ measurement methods for a single galaxy can only be applied for a handful of galaxies. Nearby early-type galaxies with signatures of regular molecular gas are ideally suited for such a cross-comparison of different methods. In this work, we have targeted the isolated (\cite[Madore et al. 2004]{Madore2004}) massive early-type galaxy NGC 6958 from the WISDOM survey (\cite[Onishi et al. 2017]{Onishi2017}). The galaxy shows clear signs of a regularly
rotating nuclear molecular gas disk and regular stellar rotation within 10$''$ from the center. As such, NGC 6958 is a good case to test and compare different
dynamical tracers.

\begin{figure}[b]
\begin{center}
 \includegraphics[width=2.5in]{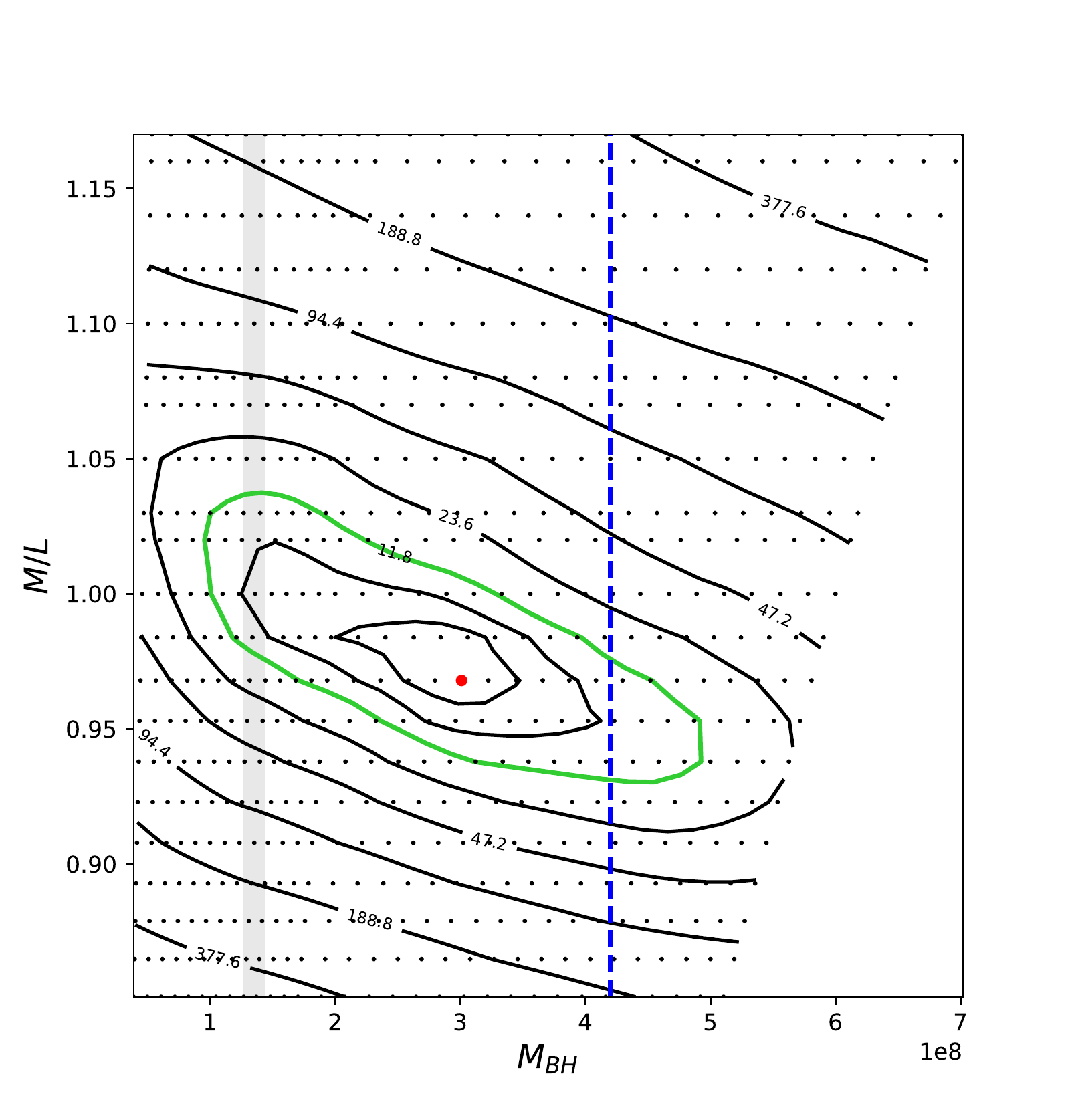}
  \includegraphics[width=2.4in]{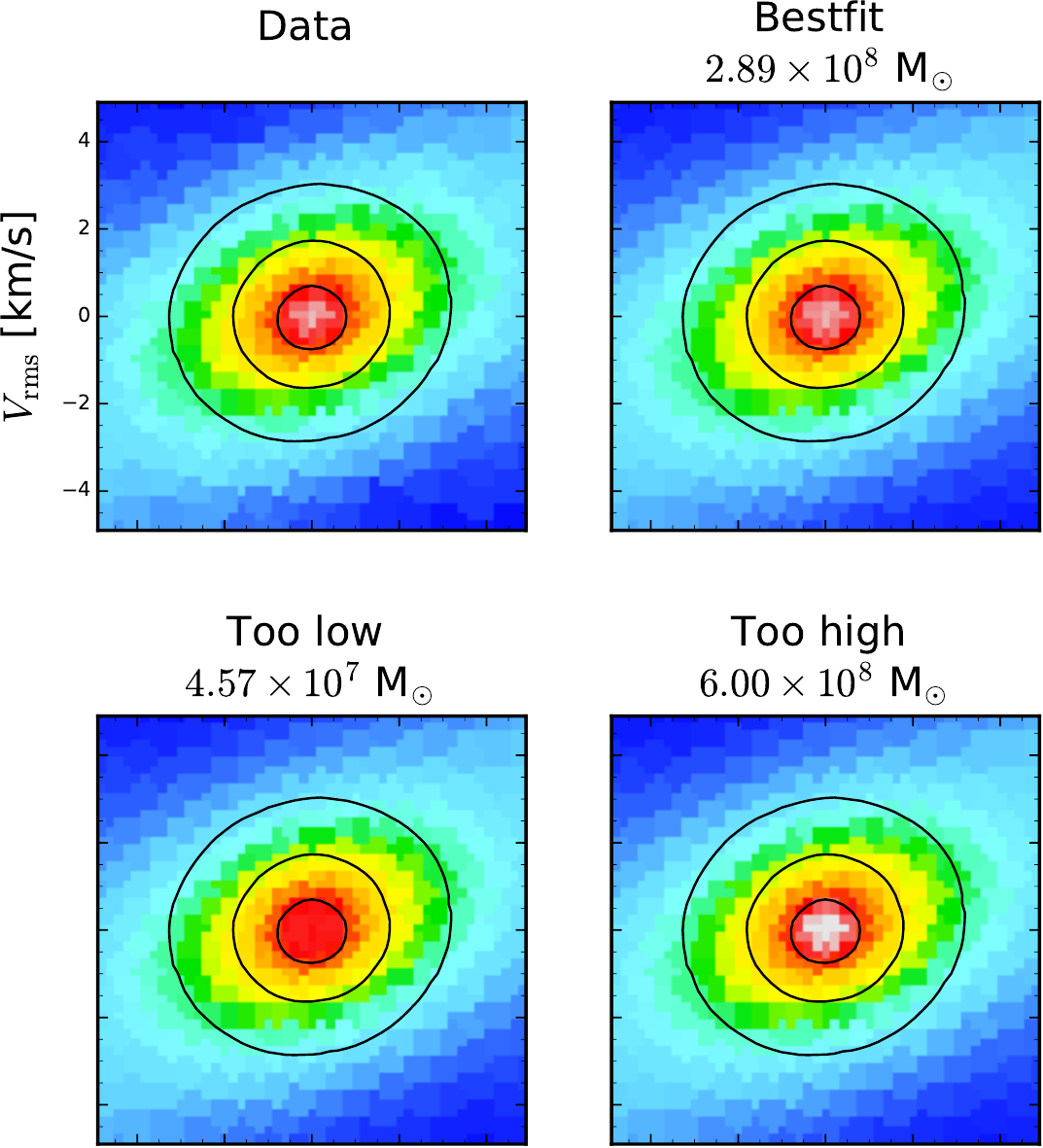}
 \caption{Results of the dynamical Schwarzschild models to estimate $M_{\rm BH}$ of NGC 6958.
  The {\it left panel} shows the grid of our models (indicated as black dots) over various M/L and $M_{\rm BH}$ values. The best-fitting model is shown
as large red dot. The overplotted contours indicate the significance, the green contour denoting
the $3\sigma$ level. We also added the $3\sigma$ threshold of the KinMS models from the molecular gas observations
(gray shaded region). The dashed blue line indicates the formally minimal $M_{\rm BH}$ that we expect to be
robustly detectable (based on the resolution of our data). The {\it right panel} shows a $V_{rms}$ comparison ([110, 250] km $s^{-1}$) between our data, the model with the best-fitting
black hole as well as models of clearly too low and too high  $M_{\rm BH}$}
   \label{fig1}
\end{center}
\end{figure}

\section{Methods}
The $M_{\rm BH}$ of NGC 6958 was estimated using two independent state-of-the-art methods based on different dynamical tracers. An overview of the methods is given in the following.

{\underline{\it Integrated stellar kinematics}}. We observed NGC 6958 as science verification project with the GALACSI adaptive-optics assisted MUSE instrument (science program 60.A-9193(A); PI= D. Krajnovi\'c). The angular resolution of our observations was determined to be 0.7$''$. From the reduced MUSE data cube, we obtained the stellar kinematics by fitting MILES (\cite[S{\'a}nchez-Blazquez et al. 2006]{Sanchez-Blazquez2006}, \cite[Falc{\'o}n-Barroso et al. 2011]{Falcon-Barroso2011}) stellar templates  to the observed spectra between 4800\,\AA\, and 6500\,\AA\, with pPXF (\cite[Cappellari \& Emsellem 2004]{Cappellari2004}). The galaxy features very regular rotation and a significant velocity dispersion within $10''$ and is therefore suitable for this test. We ran Schwarzschild (1979) orbit superposition models in a grid of constant H-band mass-to-light ratio (M/L) [0.85\,$M_{\odot}/L_{\odot}$, 1.17\,$M_{\odot}/L_{\odot}$] and $M_{\rm BH}$ [4.5e7\,$M_{\odot}$, 5.8e8\,$M_{\odot}$] at an inclination of $37^{\circ}$. In Figure 1, the grid and a comparison between the data and different models are presented. From the $\chi^2$ distribution, we derived the best-fitting parameters to be $M_{\rm BH} = (2.89 \pm 2.05 )\times 10^8\,M_{\odot}$ and $M/L = 0.98 \pm 0.06\,M_{\odot}/L_{\odot}$ within $3\sigma$ significance ($\Delta \chi^2=11.8$). For details about the method, we refer to \cite{Krajnovic2018} and \cite{Thater2019}.

{\underline{\it Molecular gas kinematics}}. We also used $^{12}$CO($2-1$) emission to trace the central gravitational potential of NGC 6958. Observations were obtained with ALMA in cycle 3 (program 2015.1.00466.S, PI: Onishi) and have a beam size of $0.7''\times 0.49''$, which is comparable to the angular resolution of the MUSE observations. We constructed a dynamical model of the molecular gas emission using the publicly available KINematic Molecular Simulation (KinMS; Davis et al. 2013), taking the beam-smearing and spatial and velocity binning into account.
Therefore, we ran a Markov chain Monte Carlo method to determine the best-fitting parameters for the black hole mass, H-band M/L, gas distribution, inclination, and additional auxiliary parameters. We ran the fitting routine for $3\times 10^6$ iterations and obtained a preliminary best-fitting black hole mass of $(1.35\pm 0.09)\times 10^8\,M_{\odot}$, mass-to-light ratio in the H-band of $0.83 \pm 0.03\,M_{\odot}/L_{\odot}$, inclination of $41.0\pm 0.2^{\circ}$ (after marginalizing over the remaining best-fitting parameters). We caution that the errors are currently very small and we will re-examine their calculation.

\begin{figure}[b]
 \vspace*{-.2 cm}
\begin{center}
 \includegraphics[width=2.4in]{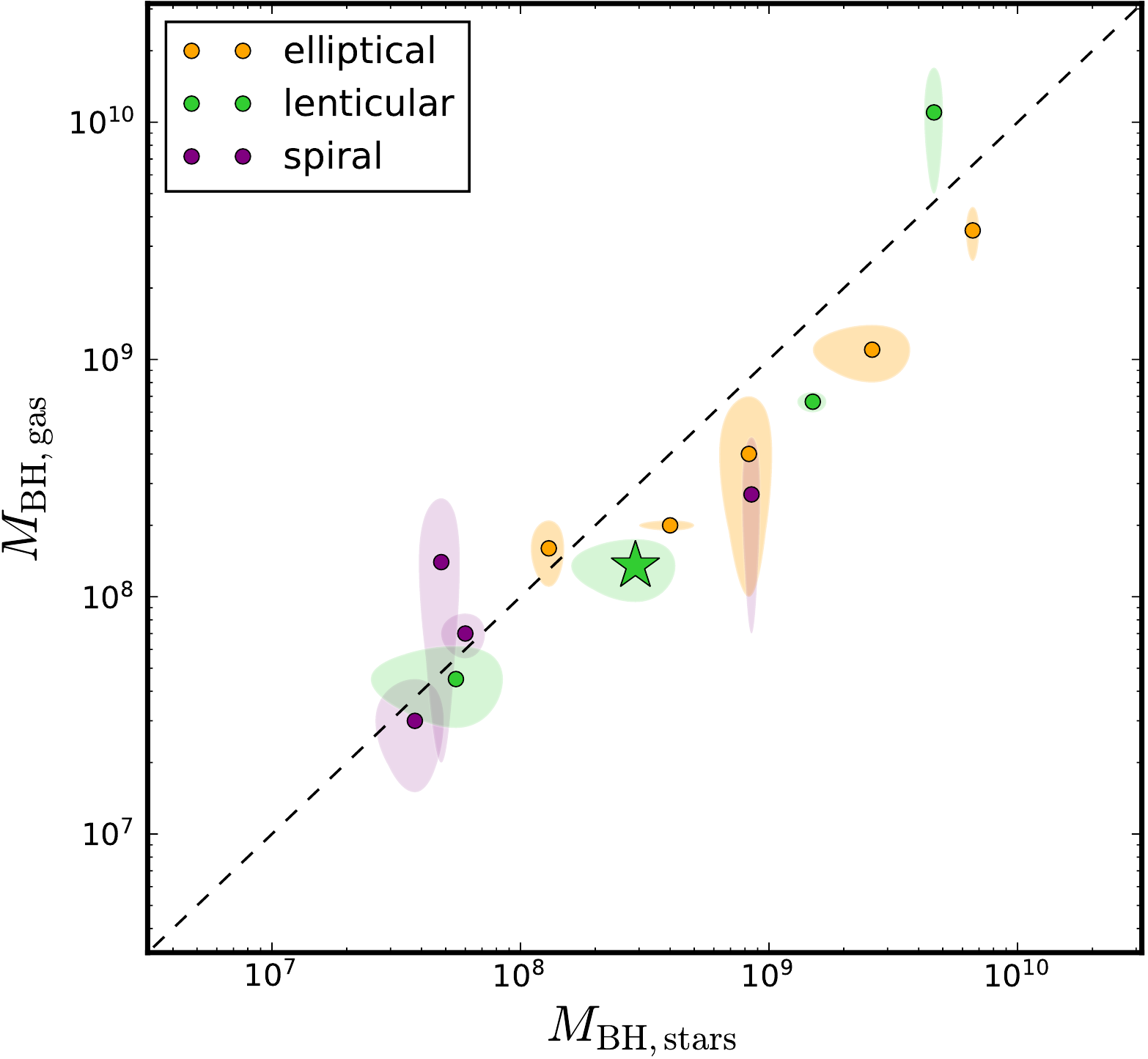}
  \includegraphics[width=2.4in]{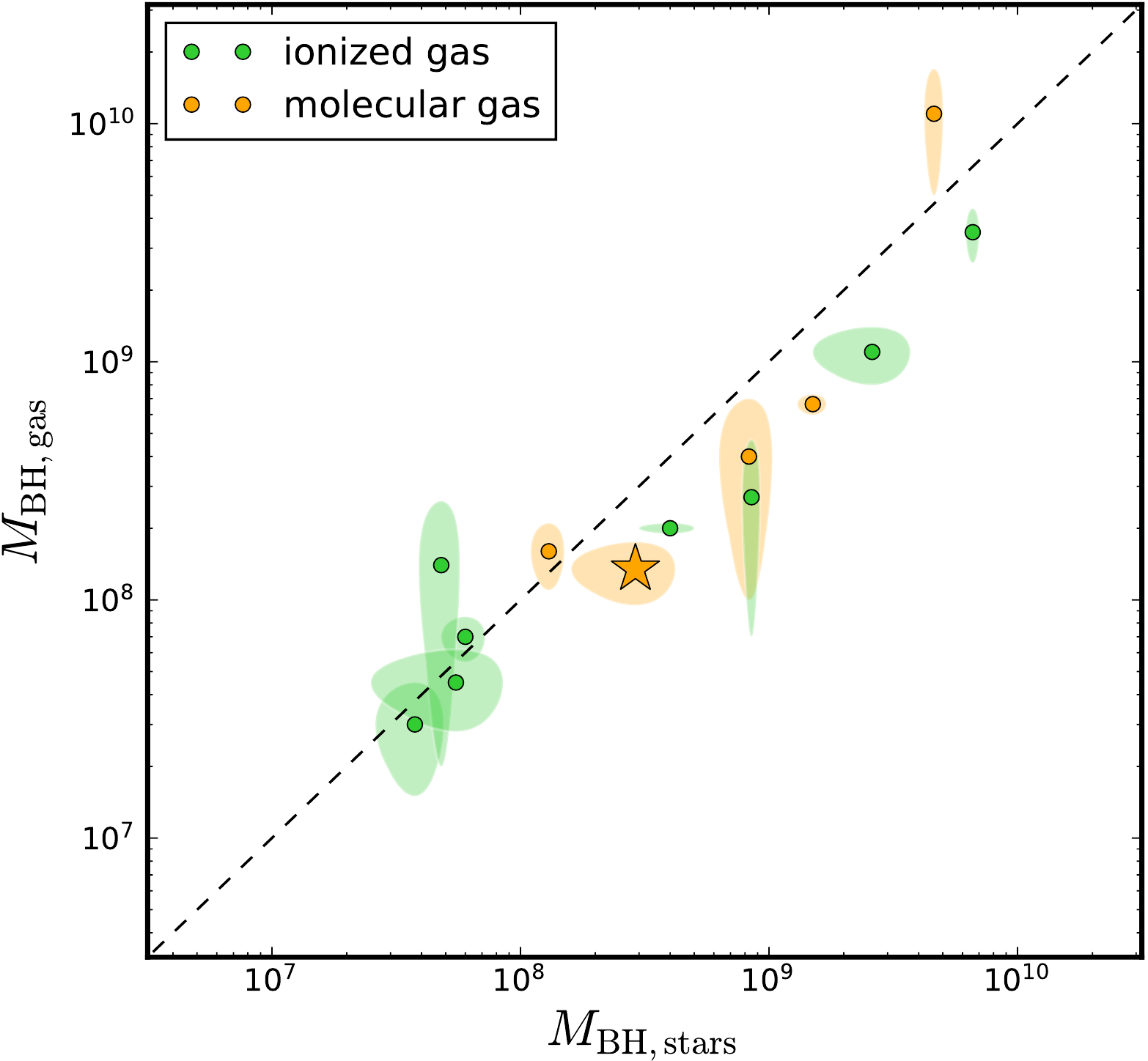}
 \caption{Cross comparisons between different $M_{\rm BH}$ measurement methods. Shown are all objects
which have both a stellar and a gas-based estimate. Colour coded are either different galaxy types ({\it left panel}) or different tracers ({\it right panel}). NGC 6948 is denoted as star. Despite the low number
statistics, gas kinematics provide significantly lower $M_{\rm BH}$ than stellar kinematics. Adapted
and extended from Kormendy \& Ho (2013).}
   \label{fig2}
\end{center}
\end{figure}

\section{Discussion and implications}
The two discussed methods provide black hole masses which are different, but consistent within their $3\sigma$ uncertainties. The derived H-band M/L is slightly higher for the stellar kinematics measurement.

{\underline{\it $M_{\rm BH}-\sigma_{\rm e}$} relation}. Using the stellar kinematic information covering the effective radius of the bulge, we also obtained the bulge effective velocity dispersion for NGC 6958 to be $\sigma_{\rm e} = 161 \pm 5$ km s$^{-1}$. The predicted value based on the M$_{\rm BH}$ - $\sigma_{\rm e}$ scaling relation is $\sim$ $1 \times 10^8\,M_{\odot}$ (\cite[van den Bosch 2016]{Bosch2016}, \cite[Saglia et al. 2016]{Saglia2016}). Both of our applied mass measurement methods provide slightly over-massive black hole compared with these scaling relations, whereas the predicted value is more consistent with the molecular gas kinematics measurement.

{\underline{\it Comparison}}. In the literature, there exists only a handful of galaxies whose black hole masses were determined through multiple methods. While these are consistent in some cases, in more than
half of the measurements there is a systematic difference between the stellar-dynamical and
gas-dynamical mass estimates (e.g., Verdoes Kleijn et al. 2002; Gebhardt
et al. 2011; Walsh et al. 2013, Barth et al. 2016), with the latter typically being significantly smaller (see
Figure 2). Our new measurements follow this trend, the stellar-based $M_{\rm BH}$ being about twice as massive as the molecular gas-based $M_{\rm BH}$. While the number statistics for the comparisons is still small, Figure 2 suggests
that the discrepancy tends to occur for elliptical galaxies, for black hole masses above $\sim$ $10^8\,M_{\odot}$ and independent of the usage of ionized or gaseous kinematics as a tracer.

{\underline{\it Systematics of the methods}}. A possible explanation is that the discrepancy between the methods is caused by inaccurate modeling assumptions in the gas dynamics method. This could be of particular concern for massive elliptical galaxies as their gas motion often suggest the presence of substantial random
motions (Noel-Storr et al. 2003, 2007) presumably agitated by nongravitational forces associated
with radio jets (Verdoes Kleijn et al. 2006). Jeter et al. (2019) analyze the effect of including
non-Keplerian motions when modeling ionized gas kinematics and conclude that the derived
masses could increase by a factor 2. Therefore, improving the models of ionized gas kinematics
could resolve the discrepancy between the ionized gas dynamics and stellar kinematics (see their
Figure 4). 
Further analysis of the systematics of the different methods is crucial to understand this mass discrepancy.

\appendix
\section{}

\begin{figure}[b]
 \vspace*{-.2 cm}
\begin{center}
 \includegraphics[width=5in]{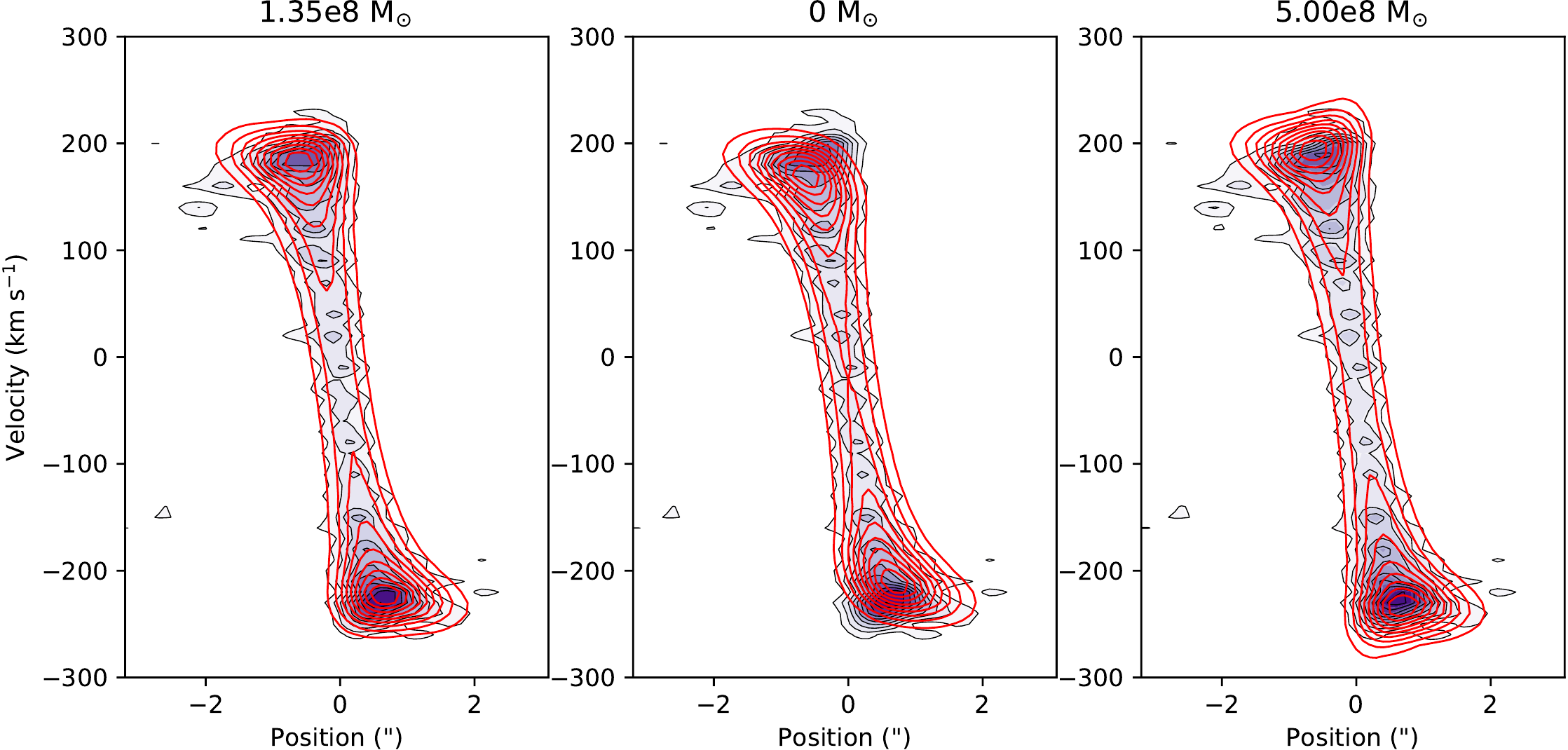}

 \caption{Position-velocity diagrams (PVD) of the $^{12}$CO(2-1) emission extracted along the major axis and overlaid with the PVD of the KinMS fit (red). Shown are the PVDs for the best-fitting model, a too low and too high $M_{\rm BH}$.  This plot is due to page restrictions not part of the original proceeding, but was added in this version to support the molecular gas kinematics result. }
   \label{fig2}
\end{center}
\end{figure}

\end{document}